\documentclass{aastex63}
\usepackage{amsmath}
\usepackage{units}
\usepackage{gensymb}

\begin{document}
\title{Exploring the stellar rotation of early-type stars in the LAMOST
Medium-Resolution Survey. II. Statistics}

\author[0000-0002-3279-0233]{Weijia Sun} 
\affiliation{Kavli Institute for Astronomy \& Astrophysics and
  Department of Astronomy, Peking University, Yi He Yuan Lu 5, Hai
  Dian District, Beijing 100871, China}
\affiliation{Key Laboratory for Optical Astronomy, National
  Astronomical Observatories, Chinese Academy of Sciences, 20A Datun
  Road, Chaoyang District, Beijing 100012, China}
  
\author[0000-0002-6573-6719]{Xiao-Wei Duan}
\affiliation{Kavli Institute for Astronomy \& Astrophysics and
  Department of Astronomy, Peking University, Yi He Yuan Lu 5, Hai
  Dian District, Beijing 100871, China}
\affiliation{Key Laboratory for Optical Astronomy, National
  Astronomical Observatories, Chinese Academy of Sciences, 20A Datun
  Road, Chaoyang District, Beijing 100012, China}

\author[0000-0001-9073-9914]{Licai Deng}
\affiliation{Key Laboratory for Optical Astronomy, National
  Astronomical Observatories, Chinese Academy of Sciences, 20A Datun
  Road, Chaoyang District, Beijing 100012, China}
\affiliation{School of Astronomy and Space Science, University of the
  Chinese Academy of Sciences, Huairou 101408, China}
\affiliation{Department of Astronomy, China West Normal University,
  Nanchong 637002, China}
  
\author[0000-0002-7203-5996]{Richard de Grijs}
\affiliation{Department of Physics and Astronomy, Macquarie
  University, Balaclava Road, Sydney, NSW 2109, Australia}
\affiliation{Research Centre for Astronomy, Astrophysics and
  Astrophotonics, Macquarie University, Balaclava Road, Sydney, NSW
  2109, Australia}
  
\begin{abstract}
Angular momentum is a key property regulating star formation and
evolution. However, the physics driving the distribution of the
stellar rotation rates of early-type main-sequence stars is as yet
poorly understood. Using our catalog of 40,034 early-type stars with
homogeneous $v\sin i$ parameters, we review the statistical properties
of their stellar rotation rates. We discuss the importance of possible
contaminants, including binaries and chemically peculiar stars. Upon
correction for projection effects and rectification of the error
distribution, we derive the distributions of our sample's equatorial
rotation velocities, which show a clear dependence on stellar
mass. Stars with masses less than $\unit[2.5]{M_\odot}$ exhibit a
unimodal distribution, with the peak velocity ratio increasing as
stellar mass increases. A bimodal rotation distribution, composed of
two branches of slowly and rapidly rotating stars, emerges for more
massive stars ($M>\unit[2.5]{M_\odot}$). For stars more massive than
$\unit[3.0]{M_\odot}$, the gap between the bifurcated branches becomes
prominent. For the first time, we find that metal-poor ([M/H] $<
-0.2$ dex) stars only exhibit a single branch of slow rotators, while
metal-rich ([M/H] $> 0.2$ dex) stars clearly show two branches. The
difference could be attributed to unexpectedly high spin-down rates and/or
in part strong magnetic fields in the metal-poor subsample.
\end{abstract}

 
\defcitealias{Sun2021}{Paper I}
\section{Introduction}
\label{sec:intro}

Stellar rotation profoundly affects almost every aspect of stellar
evolution \citep{2000ARA&A..38..143M}. It plays a crucial role in
dynamo-driven magnetic activity \citep{2002A&A...381..923S}, stellar
winds \citep{2005ApJ...632L.135M}, surface abundances
\citep{1974ARA&A..12..257P}, chemical yields
\citep{2018MNRAS.476.3432P}, internal structure
\citep{2019ARA&A..57...35A}, and external structure
\citep{2013A&ARv..21...69R}. Consequently, understanding stellar
rotation is essential for our understanding of the origin and
evolution of stellar angular momentum. However, the connection between
models and observations, although much improved in the last few
decades, still suffers from several shortcomings.

From a theoretical perspective, stellar rotation on the main sequence
(MS) is the result of the long-term evolution of angular momentum loss
and redistribution, regulated by intricate processes early in the star
formation phases, e.g. the fragmentation of rotating clouds
\citep{1979ApJ...234..289B}, magnetic braking
\citep{1985ApJ...298..190M}, bipolar overflows
\citep{1985ApJ...293..216P}, and magnetic locking with the accretion
disk \citep{2011MNRAS.416..580L}. Once the initial rotation rate at
the zero-age MS (ZAMS) has been established, the rotation evolution
can be described by a stellar rotation model that encompasses the
physical prescriptions of meridional circulation, angular momentum
transport, and (radiative, supra-Eddington, and mechanical) mass loss
\citep{2012A&A...537A.146E}. From an observational perspective, the
observed stellar rotation rates are usually entangled with projection
effects, which further complicates the use of observed rotation rates
as benchmark for stellar rotation models.

The distribution of stellar rotation rates can serve as a crucial
probe of the formation and evolution processes of stellar
populations. A sharp decline in rotation velocities of MS stars is
found around $\unit[1.2]{M_\odot}$, with more massive stars having
mean rotation velocities of hundreds of $\unit{km\,s^{-1}}$, while
lower-mass stars have velocities on the order of a few
$\unit{km\,s^{-1}}$ \citep{1970saac.book..385K}. This is mainly so,
because more massive stars lack deep convective envelopes or strong
magnetic fields that slow down stellar rotation. Many studies
\citep[e.g.,][]{1995ApJS...99..135A, 1996ApJ...463..737P,
  2006ApJ...648..580H, 2010ApJ...722..605H} have been dedicated to
deconvolve the observational distribution and derive a true rotational
distribution by means of large population statistics. A bimodal
distribution of rotation rates is observed among young solar-mass
stars in the Orion nebula \citep{1992ApJ...398L..61A,
  1999AJ....117.2941S, 2003ApJ...586..464B}. The origin of such a
bimodality is not yet fully understood, but it is thought to be a
manifestation of star--disk interactions during pre-MS evolution
\citep{1993A&A...272..176B}.

A similar bimodal distribution is also observed among massive stars
($M\gtrapprox\unit[2.5]{M_\odot}$), although the underlying mechanism
might not be the same. \citet{1995ApJS...99..135A} found a bimodal
distribution of rotation velocities among A-type MS stars in the
field, where the slowly rotating population was mainly ascribed to
chemically peculiar (CP) stars. A large-scale study by
\citet{2012A&A...537A.120Z}, who reported $v\sin i$ measurements for
2014 B6- to F2-type field stars, found a substantial difference
between the rotation distributions of stars with masses below and
above $\unit[2.5]{M_\odot}$. The less massive population exhibits a
unimodal profile, while the more massive population is bimodal. They
also revealed the complex behavior of stellar rotation, with strong
acceleration in the first one-third of the MS evolutionary phase,
followed by a mild spin-down. This calls for more efforts to improve
our understanding of stellar rotation, in terms of both theory and
observations.

Significant progress in stellar rotation statistics is now achievable
with multi-object spectrographs acquiring data for large numbers of
stars. Large spectroscopic surveys such as the William Herschel Telescope
Enhanced Area Velocity Explorer
\citep[WEAVE,][]{2012SPIE.8446E..0PD}, the Galactic Archaeology with High
Efficiency and Resolution Multi-Element Spectrograph
\citep[GALAH,][]{2021MNRAS.tmp.1259B}, the Radial Velocity Experiment
\citep[RAVE,][]{2017AJ....153...75K}, the Apache Point Observatory Galactic
Evolution Experiment \citep[APOGEE,][]{2017AJ....154...94M}, and the Large
Sky Area Multi-Object Fiber Spectroscopic Telescope
\citep[LAMOST,][]{2012RAA....12.1197C, 2020arXiv200507210L} will continue
to provide millions of stellar spectra and, hence, allow to unveil an
unprecedentedly detailed view of stellar features. In Paper I
\citep[][]{Sun2021}, we compiled a catalog containing more than 40,000
stars with effective temperatures of $\unit[7000]{K} \leq
T_\mathrm{eff} \leq \unit[14,500]{K}$ from LAMOST data release (DR) 7
medium-resolution spectra. With such a large and homogeneous sample,
we aim to review the rotation distribution of early-type stars as
function of their masses, ages, and metallicities.

This article is organized as follows. In Section~\ref{sec:data}, we
analyze the uncertainties in $v\sin i$ in more detail. The effects of
possible contamination and its influence on the derived rotation rates
are discussed in Section~\ref{sec:contamination}.
Section~\ref{sec:vdist} discusses our statistical approach to derive
the true rotational velocity distribution and we further apply this
method to the rotation rate distribution in
Section~\ref{sec:rotdist}. We also discuss the dependence of the
stellar rotation rates on metallicity, in
Section~\ref{sec:metal}. Finally, our conclusions are summarized in
Section~\ref{sec:conclusions}.

\section{Data}
\label{sec:data}

Our stellar sample was defined in \citetalias{Sun2021}, where we used
a data-driven method to estimate the stellar parameters and abundances
(`stellar labels') of early-type candidates selected from the
H$\alpha$ and Mg\,{\sc i} \textit{b} line indices provided by LAMOST
DR7 medium-resolution spectra. Having adopted a selection based mainly
on effective temperatures ($\unit[7000]{K}\leq T_\mathrm{eff} \leq
\unit[14,500]{K}$) and data quality, we derived a final catalog composed
of 40,034
stars. The precisions of our estimates are $\sim\unit[75]{K}$,
\unit[0.06]{dex}, \unit[0.05]{dex}, and $\sim\unit[3.5]{km\,s^{-1}}$
for
$T_\mathrm{eff}$, $\log g$, [M/H], and $v\sin i$, respectively, for
signal-to-noise ratios, $\mathrm{SNR} > 60$. However, this only
represents the average behavior of our estimates; the actual
performance may vary \citep{Sun2021}. As regards the observational
uncertainties, they may originate from the line profiles of a single
exposure and/or from minor variations among multiple-epoch
observations.

\begin{figure}[ht!]
\plotone{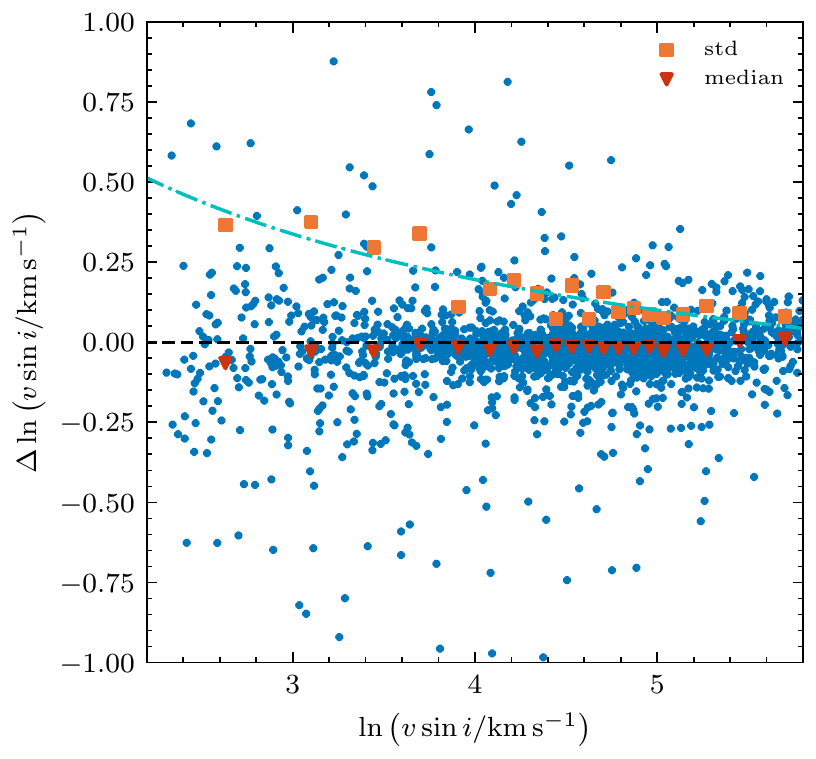}
\caption{Variations of $v\sin i$ estimates from single-epoch
  observations with respect to the mean value. The data was analyzed
  using constant numbers in each bin. Orange squares represent the
  1$\sigma$ uncertainties, $\Delta\ln v\sin i$, and red triangles
  represent the median value (bias). An inverse proportional function
  (cyan dot-dashed line) is overplotted to fit the standard
  deviations. \label{fig:error}}
\end{figure}

To investigate the dependence of the uncertainty on $v\sin i$, we used
the difference between the individual $v\sin i$ values from the
single-epoch observations and the associated mean value for the
co-added spectra. There ought to be only minor differences among the
rotation velocities of single-epoch exposures; their difference
represents the uncertainty in our
estimates. \citet{2002A&A...381..105R} found that the error in
$\ln(v\sin i)$ can be approximated by a normal distribution, based on
multi-epoch (48), high-SNR ($>250$) observations of Sirius. This is
because fast rotation blends the absorption lines and affects the
accuracy of radial and rotation velocity measurements. Therefore, the
variation was calculated in logarithmic space (see
Figure~\ref{fig:error}). For $\ln(v\sin i) < 4$ ($v\sin i <
\unit[50]{km\,s^{-1}}$), the standard deviation (orange squares)
exhibits a dependence on the $v\sin i$ value. This is at odds with the
result of \citet{2002A&A...393..897R, 2007A&A...463..671R}, who found
a uniform error value on a logarithmic scale. This is due to the
differences between the methods used to estimate rotation velocities.

Because those earlier authors adopted the first zero of the Fourier
transform of the line profiles \citep{1933MNRAS..93..478C,
  1976oasp.book.....G} based on high-resolution spectra ($R > 20,000$)
to determine $v\sin i$, their estimates are less affected by
resolution limitations. However, our data from the LAMOST MRS are
unavoidably further broadened due to the finite resolution power of
the instrument, which is known as instrumental broadening. The
observed line profiles are a convolution of the true profiles and the
instrumental profile, which makes it hard to estimate very slow
rotation velocities and, hence, results in larger uncertainties for
slow compared with fast rotators.

Based on our estimates using a spectrograph with $R\sim 4000$
\citep{2019ApJ...876..113S}, the detection limit for the MRS data
($R\sim 7500$) is around $\unit[30]{km\,s^{-1}}$. For these slow
rotators, the relative uncertainties are comparable to the absolute
values, and so such data should be used with caution. We used an
inverse proportional function to fit the standard deviations. The
median values of $\Delta \ln v\sin i$ are also presented as red
triangles in Figure~\ref{fig:error}. For stars with rotation
velocities above the detection limit, the relative
bias---$\mathrm{Bias}(v\sin i)/v\sin i$---is smaller than 3\%.

\section{Contamination}
\label{sec:contamination}

The rotation behavior of our sample stars might not be solely the
result of their own evolution but it could also be affected by several
external processes. Therefore, possible polluters need to be discarded
first. In this section, we present two possible contaminants: binary
and CP stars. We check for the incidence of periodic variables and
star cluster members in Section~\ref{sec:var} and
Section~\ref{sec:cluster}, respectively. We then discuss the different
rotation properties of these contaminants and the effects of possibly
having missed contaminants on the rotation distribution of `normal'
(single, non-CP, non-variable, non-cluster member) stars. Following
removal of all identifiable contaminants, we are left with 33,655
normal stars (see Section 4.2).

Binaries, particularly close binaries, can drastically affect the
rotation properties of both of their components through tides, mass
transfer, or even mergers \citep{2013ApJ...764..166D}. Tidal
interactions tend to synchronize stellar and orbital rotation,
transferring the stellar spin angular momentum to the orbital angular
momentum of the binary system. Mass transfer from the donor star can
spin up the accreting star and cause it to approach the break-up
limit. Mergers may also change stellar rotation, although this is the
least well-understood process.

CP stars represent another possible contaminant type in our
sample. They are early-type MS stars exhibiting anomalous chemical
abundances. According to \citet{1974ARA&A..12..257P}'s classification,
they can be divided into four subgroups: metallic line (Am),
magnetically peculiar (mAp), stars with enhanced Hg {\sc ii} and Mn
{\sc ii} (HgMn), and He-weak stars. Most CP stars exhibit slow
rotation, but the distribution may differ among
subgroups. \citet{1995ApJS...99..135A} found that virtually all Am and
mAp stars have equatorial rotation velocities below
$\unit[120]{km\,s^{-1}}$.

\subsection{Contamination by binary stars}
\label{sec:binary}

If a binary system's stellar rotation is modified by interactions
between the components, that should also affect the system's radial
velocity, $v_{R}$. Variations in the latter could be identified
through Doppler shifts in the case of spectroscopic binaries. Ideally,
such shifts should be periodic, following the orbital period, which
could be confirmed by means of multi-epoch, high-cadence
observations. However, only 1\% of our sample objects have been
observed more than 10 times, while most ($>60$\%) only have three
epochs of observations. Therefore, we used the variations in $v_{R}$
between single-exposure spectra to tentatively infer binarity.

Relative variations were estimated following Xiong et al. (2021, in
prep.). We relied on relative variations to avoid systematic errors in
the absolute $v_{R}$ measures \citep{2018AJ....156...90H}. In brief,
we adopted the single-exposure spectrum with the highest SNR as our
spectral template and calculated $\Delta v_{R}$ for the other spectra
through cross-correlation. To differentiate binaries from single
stars, we introduced $\sigma_\mathrm{detect}$
\citep{2013A&A...550A.107S}, i.e. the ratio of the relative $\Delta
v_{R}$ to its uncertainty. We considered a star to be a binary system
if any pair of single-exposure spectra met the following condition:
\begin{equation}
\sigma_\mathrm{detect} = \frac{|v_i-v_j|}{\sqrt{\sigma^2_i+\sigma^2_j}} >
C_1\quad \mathrm{and}\quad |v_i-v_j| > C_2,
\end{equation}
where $|v_i-v_j|$ is the maximum difference between these
measurements. The latter criterion was introduced to exclude fake
binary stars, i.e. stars with intrinsic variability that may be caused
by photospheric or wind effects.

We adopted ($C_1 = 4$, $C_2 = 16$) as our selection criteria for
binary systems. We detected 4955 binaries from a sample of 36,620
candidates ($N_\mathrm{obs} \geqslant 3$). The corresponding binary
fraction for this sample is $(13.5\pm 0.3)$\%. This is lower than the
binary fraction observed for O-type \citep{2013A&A...550A.107S} and
early-B type stars \citep{2015A&A...580A..93D, 2017IAUS..329..110S},
but consistent with \citet{2021arXiv210909775G}'s result for late-B and early-A
type stars. Note that a large fraction of binaries will remain
undetected. \citet{2021arXiv210909775G} used a Monto Carlo method to correct for
observational biases and estimated that the intrinsic binary fraction
for our sample could be as high as about 40\%. This type of analysis
is not necessary for the context in which we are looking for
contaminants. Our method to detect spectroscopic binaries is most
sensitive to close binaries that could influence the stellar rotation
rates of their components. Any missing binaries are mostly long-period
systems, which are not expected to affect the rotation rates of their
individual components.

\subsection{Contamination by CP Stars}

Identification of CP stars relies on strong (or weak) absorption lines
of certain elements in their spectra. However, this is currently not
feasible for our sample. First, the wavelength coverage of the LAMOST
MRS is rather narrow and misses important lines that are used to
identify anomalies (e.g., the Ca\,{\sc ii} H and K lines). However, a
detailed analysis of the chemical abundances of early-type stars based
on our MRS spectra is beyond the scope of this study, and we defer the
identification of CP stars using the same data to future papers.

Instead, we used the pre-compiled catalogs of
\citet{2019ApJS..242...13Q}, \citet{2020A&A...640A..40H}, and
\citet{2021A&A...645A..34P} to identify Am, mAp, and HgMn stars (the
most frequent CP subgroups) in our sample. These three papers are all
based on previous LAMOST DRs, which hence enables us to better
estimate the frequency of CP-star
occurrence. \citet{2019ApJS..242...13Q} used a random forest
machine-learning algorithm trained on the sample from the study of
\citet{2015MNRAS.449.1401H} to search for Am stars from LAMOST DR5
high-SNR ($> 50$) early-type LRS spectra and they also used the
equivalent width of the \unit[4077]{\AA} line, a blended line of
Sr\,{\sc ii}, Cr\,{\sc ii}, and Si\,{\sc ii}, to label mAp
candidates. We reselected our sample following their selection
criteria, and cross-matched our sample with their catalog. We found
405 Am stars and 24 mAp candidates in a subsample of 8051
objects. \citet{2020A&A...640A..40H} and \citet{2021A&A...645A..34P}
applied {\bf a different approach based on MKCLASS to their DR4-based sample. We obtained 46
mAp and 12 HgMn stars among 7676 candidates by cross-matching the DR4 data with their catalogs}. Note that the selection
of the subsamples in the latter cases is not the same as that
described by \citet{2020A&A...640A..40H} and
\citet{2021A&A...645A..34P}, who used \textit{Gaia} DR2 colors, as
well as spectral types, to perform early-type sample selection. We also
cross-matched our sample with \citet{2009A&A...498..961R}, a catalog
compiled from the literature. Ultimately, we identified 522 Am, 115
mAp, 13 HgMn, and 1 He-weak stars.

The apparent detection rate of Am stars in our subsample is around
5\%, which is similar to that reported by
\citet{2019ApJS..242...13Q}. However, given that Am stars are mostly
found around $T_\mathrm{eff}\sim\unit[7750]{K}$, the detection rate based on a
sample with a wide range of temparture does not properly reflect the
incidence. In the most likely temperature range,
the frequency of
occurrence is more than 15\%, but it is still lower than the values
reported in previous studies \citep{1974ARA&A..12..257P,
  1981ApJS...45..437A, 2012A&A...537A.120Z}. This is probably due to
\citet{2019ApJS..242...13Q}'s strict screening, since they conducted
additional manual identification based on spectral features associated
with chemical peculiarity.

\subsection{Contamination by periodic variables}
\label{sec:var}

Our comparison catalog of periodic variables was recently published by
\citet{2020ApJS..249...18C}. The authors searched for and classified
781,602 variables into 11 main types based on the Zwicky Transient
Facility \citep[ZTF;][]{2019PASP..131f8003B} DR2, which covers the
northern sky. We also cross-matched our sample with a combination of
other catalogs, including the Wide-field Infrared Survey Explorer
\citep[WISE,][]{2018ApJS..237...28C}, the Asteroid Terrestrial-impact Last
Alert System
\citep[ATLAS,][]{2018AJ....156..241H}, the All-Sky Automated Survey for
Supernovae
\citep[ASAS-SN,][]{2018MNRAS.477.3145J}, and the Catalina survey
\citep{2014ApJS..213....9D}. We found that our sample
contained 325 periodic variables.

\subsection{Contamination by cluster members}
\label{sec:cluster}

One of the LAMOST-II MRS survey's science goals pertains to open
clusters. The survey's target sample includes 18 open cluster-related
fields, containing several tens of open clusters
\citep{2020arXiv200507210L}. It has been reported that the
distribution of the rotation velocities of cluster members might be
different from that of field stars \citep{2006ApJ...648..580H}, in the
sense that cluster B-type stars contain fewer slow rotators relative
to field B-type stars. Although \citet{2010ApJ...722..605H} argued
that the lower rotation rates of the field stars are primarily the
result of evolutionary spin-down changes, we removed any cluster
members to retain a clean sample of early-type field stars. To do so,
we adopted the cluster member catalog of \citet{2020A&A...633A..99C},
who applied a membership assignment procedure to the \textit{Gaia} DR2
dataset and identified members of 1481 clusters. We found that our
sample included 726 cluster members, equivalent to less than 2\% of
the total sample.

\subsection{$v\sin i$ contamination}
\label{sec:vsini_cont}

Even though we have tried to filter out most contaminants, a fraction
may have been missed by our decontamination method, while the (poor)
decontamination completeness may compromise estimates of the rotation
distribution. It is, therefore, necessary to investigate the rotation
properties of the contaminants discussed in the previous sections.

\begin{figure}[ht!]
\plotone{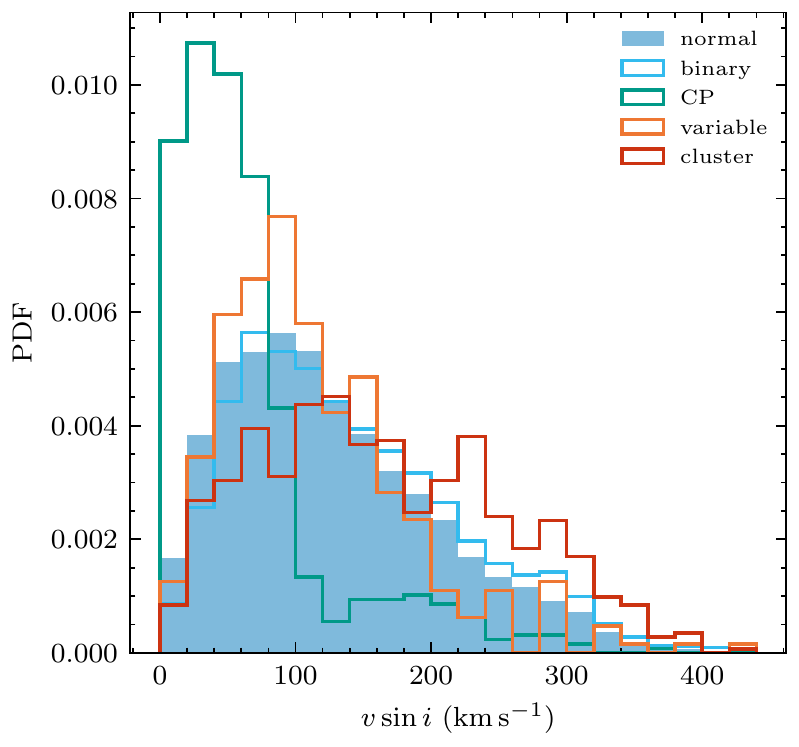}
\caption{$v\sin i$ distributions of (blue) normal stars, (cyan) binary
  stars, (green) CP stars, (orange) periodic variables, and (red)
  cluster members. \label{fig:vsini_contamination}}
\end{figure}

In Figure~\ref{fig:vsini_contamination}, we present $v\sin i$
histograms for the normal stars as well as the contaminants. Only CP
stars (green) exhibit a significant deviation, in the sense that they
are mostly slow rotators. The other objects display similar profiles
as the normal stars (note that cluster members may exhibit a slightly
different $v\sin i$ distribution). As long as the $v\sin i$
contamination is similarly distributed as the $v\sin i$ distribution
of the normal stars, the overall rotation distribution will not be
affected even if a fraction of contaminants remains unidentified in
our final sample. As for the cluster subpopulation, we note that the
completeness of this catalog should be close to unity since it was
derived using high-precision proper motion data from \textit{Gaia}.

\begin{figure*}[ht!]
\plotone{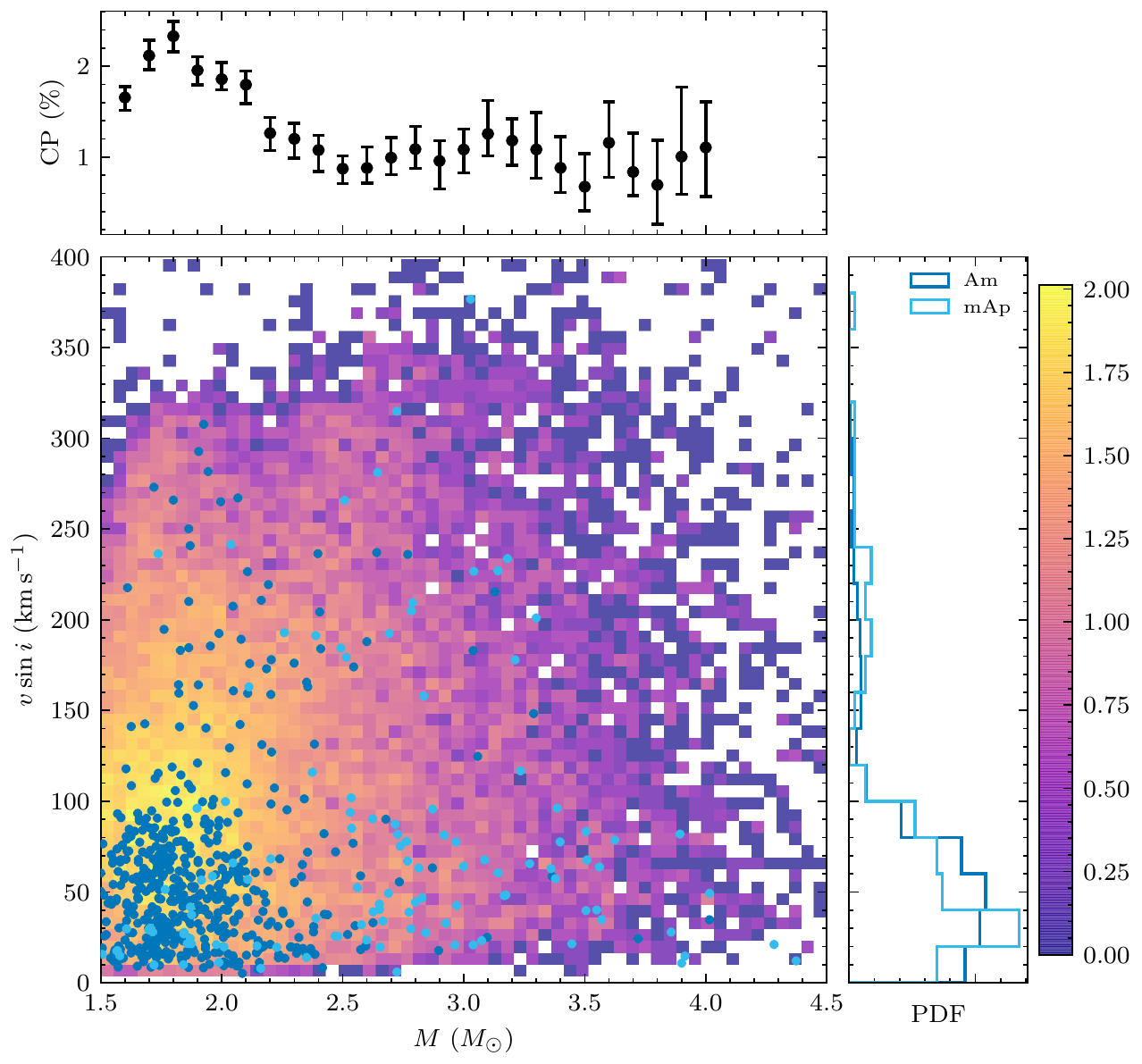}
\caption{$v\sin i$ as a function of stellar mass, $M$, for CP
  stars. The corresponding distribution of normal stars is presented
  in the background. Am and mAp stars are represented by blue and cyan
  circles, respectively. The right panel shows histograms of the
  $v\sin i$ distribution of these two subpopulations. The top panel
  shows the number fraction of CP stars in each mass bin, from
  $\unit[1.6]{M_\odot}$ to $\unit[4.0]{M_\odot}$ with a bin width of
  $\unit[0.2]{M_\odot}$. \label{fig:cp}}
\end{figure*}

CP stars tend to be slow rotators \citep[e.g.,][]{1974ARA&A..12..257P,
  1995ApJS...99..135A}. Chemical peculiarities in Am and mAp
  stars are presumed to have been caused by gravitational settling and
  radiative levitation, which can occur in stable atmospheres
  associated with slow rotation rates. However, the existence of slow
  rotators but without chemical anomalies challenges the connection
  between their chemical peculiarities and stellar rotation. We
confirm their special rotation properties in Figure~\ref{fig:cp},
where it is clear that most of CP stars have small $v\sin i$ values,
which is rather different from the behavior of the normal stars in the
$M$ versus $v\sin i$ diagram. Since most of the CP stars detected are
Am and mAp stars, only these two subtypes are shown. Am stars (blue)
span a narrow mass range from $\unit[1.5]{M_\odot}$ to
$\unit[2.0]{M_\odot}$; they are less massive than mAp stars. This is
consistent with the classification scheme of CP subtypes. In the right
panel, we present the normalized probability density function (PDF) of
the $v\sin i$ values of Am and mAp stars. We note that 10--20\% of
these CP stars have relatively large rotation velocities ($v\sin i >
\unit[120]{km\,s^{-1}}$), which was also noted by
\citet{2017MNRAS.468.2745N}. The presence of such rapidly rotating CP
stars, if confirmed, would shed new light on the possible role of
stellar rotation in the context of CP star formation
\citep{2009AJ....138...28A}.

The top panel of Figure~\ref{fig:cp} shows the number fraction of CP
stars ($N_\mathrm{CP}/N_\mathrm{total}$) in each mass bin from
$\unit[1.6]{M_\odot}$ to $\unit[4.0]{M_\odot}$ with a bin width of
$\unit[0.2]{M_\odot}$. The average CP fraction is around 1--2\% across
the mass range, and any differences in CP fraction may be attributed
to the various subtypes of CP stars populating the different mass
domains. However, based on estimates of the frequency of CP stars by
\citet{1974ARA&A..12..257P}, \citet{1981ApJS...45..437A}, and
\citet{2012A&A...537A.120Z}, 15--30\% of solar neighborhood stars (
covering
$\unit[7000]{K} < T_\mathrm{eff} < \unit[7800]{K}$) are CP stars. It thus
appears that a fraction of missing
CP stars may remain undetected in our sample, and they may bias the
rotation distribution of the normal stars. However, statistical
corrections are difficult and imprecise, because the observed
incidence may depend on the temperature as well as the detection
method.

\section{Distributions of equatorial rotation velocities}
\label{sec:vdist}

To obtain the equatorial rotation velocities, $v$, we need to correct
the projected rotation velocities $v\sin i$ for the effects of
inclination, $i$. For individual stars, this is only feasible in rare
cases, e.g. for eclipsing binaries (where strong tidal effects ensure
that the rotation axis is parallel to the orbital axis) and for stars
with a rotational modulation signal in their light curves, among
others.  However, we can correct for projection effects statistically
under the assumption of randomly oriented rotation axes
\citep{1995ApJS...99..135A, 2007A&A...463..671R, 2012A&A...537A.120Z}.

To estimate the distribution of the equatorial rotation velocities, we
followed \citet{2007A&A...463..671R} and \citet{2012A&A...537A.120Z},
whose method can be summarized in three steps:

\begin{enumerate}
\item We first smooth the distribution using a Gaussian kernel. The
  selection of the bandwidth of the kernel-smoothing window is done
using Silverman's rule of thumb \citep{1986desd.book.....S} which minimises
the
mean integrated squared error
  \begin{equation}
  	h = \left(\frac{4\hat{\sigma}}{3n}\right)^{\frac{1}{5}}
  \end{equation}
where $h$ is the bandwidth of the smoothing parameter, $\hat{\sigma}$ is the
standard deviation of the samples, and $n$ is the sample size.

\item We then recover the PDF of the true projected rotation
  velocities, $\theta \equiv \widehat{v\sin i}$, through rectification
  of the error law. As our measurements are subject to random
  uncertainties, the observed $v\sin i$ represents the maximum of the
  probability distribution, whereas $\sigma_{v\sin i}$ is the width of
  the distribution. \citet{2007A&A...463..671R} assumed that the error
  in $\ln(v\sin i)$ was normally distributed, with a mean of 0 and a
  variance of $\left(\ln(1+\alpha)\right)^2$, where the coefficient
  $\alpha = 0.1$ is a constant. However, as shown in
  Figure~\ref{fig:error}, our estimates of $v\sin i$ are susceptible
  to instrumental broadening for small $v\sin i$. Therefore, $\alpha$
  depends on the value of $v\sin i$, which must hence be implemented
  into our calculation. The deconvolution is carried out using the
  \citet{1974AJ.....79..745L} iterative technique.

Although our $v\sin i$ estimates are consistent with literature values
up to $\sim\unit[400]{km\,s^{-1}}$, they could be affected by
temperature and gravity inhomogeneities introduced by rapid
rotation. Gravity darkening suppresses the equatorial region's
contribution to the rotational broadening, which has been discussed
extensively in the context of Be stars, i.e. B-type stars exhibiting
Balmer-line emission \citep[e.g.][]{2004MNRAS.350..189T}. Nonetheless,
a correction for this underestimation is only necessary if the rapid
rotators represent a
statistically significant fraction \citep{2012A&A...537A.120Z}.

\item Following correction for the error law, the final step is to
  correct the distribution for inclination effects. Under the
  assumption of a random distribution of rotation axes, the PDF of the
  corrected projected rotation velocities, $\Psi(\theta)$, becomes
\begin{equation}
\Psi(\theta) = \int \gamma(v)P(\theta|v) \mathrm{d} v =  \int \gamma(v)
\frac{\theta}{v}\frac{H(v-\theta)}{\sqrt{v^2-\theta^2}} \mathrm{d}v,
\end{equation}
where $\gamma(v)$ is the true equatorial rotation velocity
distribution and $P(\theta|v)$ is the conditional probability. $H(x)$
represents the Heaviside step function, which is 1 if $x\geq 1$ and
otherwise 0.
\end{enumerate}

\begin{figure*}[ht!]
\plottwo{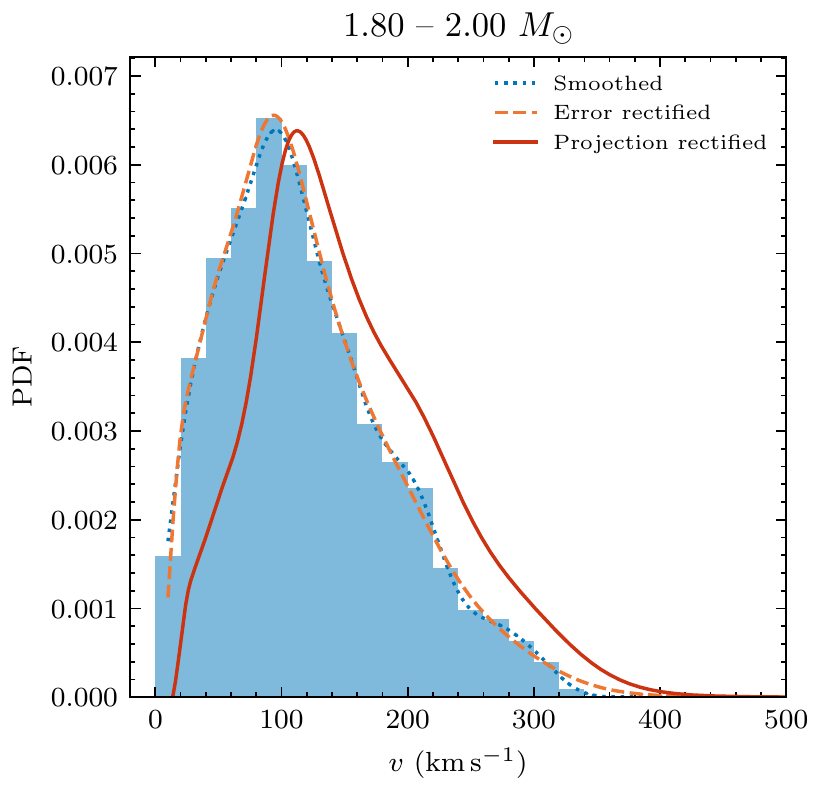}{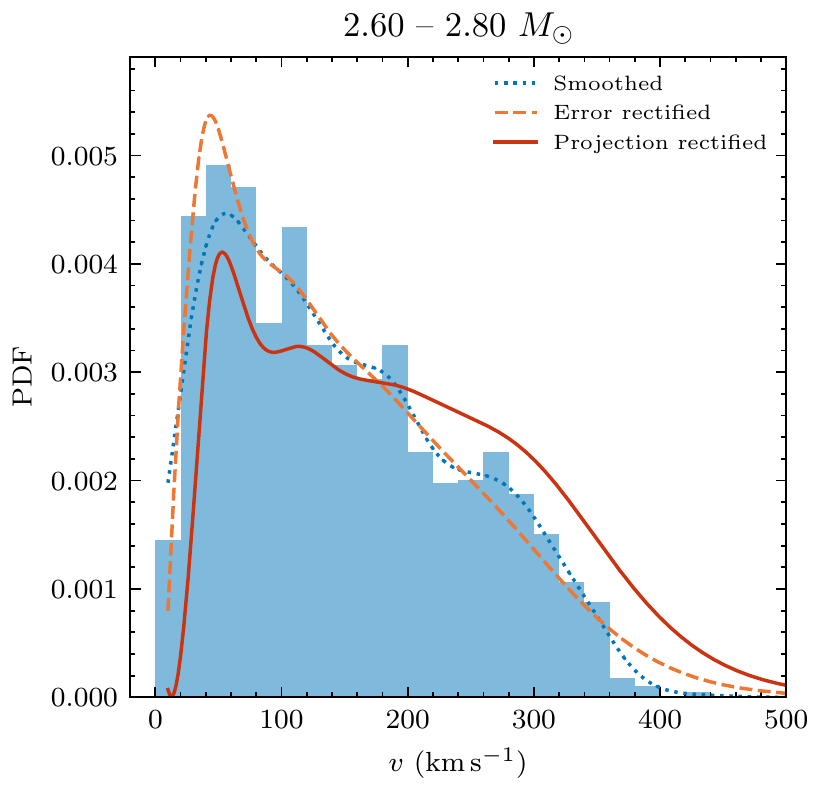}
\caption{Distributions of rotation velocities of normal stars in
  different mass bins. (left) 1.8--$\unit[2.0]{M_\odot}$; (right)
  2.6--$\unit[2.8]{M_\odot}$. The cyan histograms represent the
  observed $v\sin i$ distribution using a bin width of
  $\unit[20]{km\,s^{-1}}$, while the blue dotted lines are the
  corresponding smoothed distributions. The error-rectified
  distributions are shown as the orange dashed lines. The final
  products---projection-corrected equatorial velocity
  distributions---are shown as the red solid
  lines. \label{fig:vexample}}
\end{figure*}

We provide two examples for different mass ranges (from 1.8 to
$\unit[2.0]{M_\odot}$ and from 2.6 to $\unit[2.8]{M_\odot}$) to
illustrate the process of deconvolution: see
Figure~\ref{fig:vexample}. The original observed distribution of
$v\sin i$ is shown as the cyan histograms with a bin width of
$\unit[20]{km\,s^{-1}}$. Its smoothed $v\sin i$ distribution,
error-corrected true projected rotation velocity ($\theta$)
distribution, and projection-rectified true equatorial rotation
velocity ($v$) distribution are represented by the blue dotted line,
the orange dashed line, and the red solid line, respectively.

These two mass ranges were selected to represent the typical situation
in the less massive ($M<\unit[2.5]{M_\odot}$) and more massive
($M>\unit[2.5]{M_\odot}$) regimes. \citet{2012A&A...537A.120Z} found
that less massive stars have a unimodal equatorial velocity
distribution, while more massive stars trace a bimodal
distribution. Our results are largely consistent with previous
results, except for several significant differences:

\begin{itemize}
\item The velocity distribution of the less massive stars
  (Figure~\ref{fig:vexample}, left) shows a peak velocity at
  $v\approx\unit[110]{km\,s^{-1}}$, which is
  $\sim\unit[40]{km\,s^{-1}}$ smaller than the peak velocity reported
  by \citet{2012A&A...537A.120Z} for the same mass range.

\item For the more massive stars, we found an excess of stars with
  $v\sin i < \unit[100]{km\,s^{-1}}$, which results in a broad and
  rather flat rectified distribution. This is different from the
  bimodal distribution found by \citet{1995ApJS...99..135A} and
  \citet{2012A&A...537A.120Z}. The distribution in the right panel of
  Figure~\ref{fig:vexample} could be interpreted as showing two
  populations, with one group representing slow rotators that share
  similar rotation velocities, and a second group comprising fast
  rotators whose velocities are widely spread. If this were correct,
  we could still claim a bimodal distribution for our more massive
  sample. Note that the peak velocity of the slow rotators is similar
  to that found by \citet{2012A&A...537A.120Z}. However, the number
  fraction of slow rotators is rather distinctive in both
  papers. \citet{2012A&A...537A.120Z} found that the maximum PDF for
  slow rotators is smaller than that for fast rotators, and the number
  fraction for the former population is less than 20\%. However, slow
  rotators provide a more distinct contribution to our sample.

\item \citet{2012A&A...537A.120Z} claimed that the sum of two
  Maxwellian functions can fit the rectified distributions.  They used
  the derived Maxwellian functions to quantify the typical velocities
  and number fractions of slow and fast rotators. This was based on
  \citet{1967mrs..conf..181D}, who showed that the distribution of $v$
  can be described by Maxwellian functions. However, we note that this
  practice may not always yield a reliable representation of the $v$
  distribution, especially for the rapidly rotating population.
\end{itemize}

We suggest that the difference in peak velocity for the less massive
stars and the excess of slow rotators among more massive stars can be
attributed to missed contamination by CP stars. As discussed in
Section~\ref{sec:vsini_cont}, CP stars are mainly slow rotators and we
expected to have $\sim15$\% of such stars misclassified as normal
stars. Given these missing CP stars, the peak velocity of the low-mass
stars could be biased toward the lower end and a possibly bimodal
distribution could be hidden. This may also explain an excess of slow
rotators compared with previous studies.

However, recent papers have suggested the existence of a population of
slowly rotating normal early-type stars \citep{2020MNRAS.498.2456S,
  2021arXiv210504741Q}. Most of these studies are based on photometric
variability caused by the inhomogeneous structure of the stellar
surface, which could be used to infer the rotation
period. \citet{2021arXiv210504741Q} searched for low-velocity rotating
normal (non-CP) A-type stars from the LAMOST--\textit{Kepler} project
and found a significant fraction of slow rotators among normal A-type
stars. Such a contradiction is also reported for the 30 Doradus
starburst region in the Large Magellanic Cloud (LMC). The O-type stellar
sample \citep{2013A&A...560A..29R} shows a similar rotation
distribution as the right panel of Figure~\ref{fig:vexample}),
exhibiting an excess of slow rotators, while the B-type stellar sample
\citep{2013A&A...550A.109D} present a bimodal population of very slow
and fast rotators, with few stars rotating at the low-velocity peak
seen for the O-type stars. These controversies may suggest that
previous studies of the rotation properties of early-type stars may
have missed a significant number of slow rotators.

\section{Distributions of rotation rates}
\label{sec:rotdist}

For a single star that is not affected by binary interactions, we can
generalize the formalism of the observed rotation velocity, $v$, by
writing out its dependence on mass and stellar age:
\begin{equation}
	v(M, \omega_\mathrm{init}, t/t_\mathrm{MS}),
\end{equation}
where the stellar age is expressed as the relative age and the initial
rotation rate is given by
$\omega_\mathrm{init}=\Omega/\Omega_\mathrm{crit}$
\citep{2013A&A...553A..24G}. The parameter $\Omega$ is the angular
velocity and $\Omega_\mathrm{crit}$ is the critical velocity where the
surface gravity can no longer maintain equilibrium with the
centrifugal motion. Since $v$ changes as a function of time, and also
depends on the stellar mass, the parameter $v$ itself is not suitable
for comparison among samples with different properties.

\begin{figure}[ht!]
\plotone{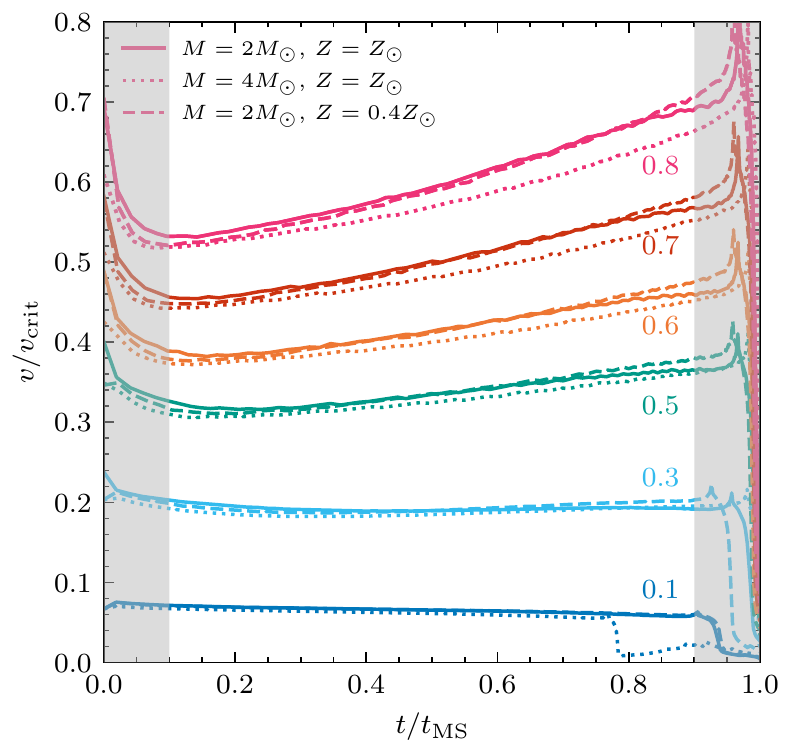}
\caption{$v/v_\mathrm{crit}$ evolution \citep{2013A&A...553A..24G} for
  different initial rotation rates (0.1, 0.3, 0.5, 0.6, 0.7,
  0.8). Solid lines represent a $\unit[2]{M_\odot}$ star of solar
  abundance, dotted lines represent a $\unit[4]{M_\odot}$ star of
  solar abundance, while dashed lines represent a $\unit[2]{M_\odot}$
  star of $\unit[0.4]{Z_\odot}$. The region where $v/v_\mathrm{crit}$
  shows a strong dependence on relative age is marked in gray. For
  $0.1 \leq t/t_\mathrm{MS} \leq 0.9$, $v/v_\mathrm{crit}$ could be
  used as a proxy for the initial rotation rate. \label{fig:omega}}
\end{figure}

We suggest this dependence on mass and age can be alleviated by
introducing $v/v_\mathrm{crit}$
\citep[e.g.][]{2017MNRAS.468.2745N}. For slow rotators, this ratio is
almost constant, particularly for rigid rotators on the MS
\citep{2012A&A...537A.120Z}. We illustrate this by exploring the
evolutionary models of the Geneva group \citep{2012A&A...537A.146E,
  2013A&A...553A..24G}. In Figure~\ref{fig:omega}, we show the
$v/v_\mathrm{crit}$ evolution for different initial rotation
rates. The ratio is almost constant for $0.1 \leq t/t_\mathrm{MS} \leq
0.9$ if $\omega < 0.5$. For fast rotators, the ratio exhibits a mild
increase with time. This is to account for the enhanced `mechanical'
mass loss in the equatorial region as a star reaches the critical
velocity. Even though the $v/v_\mathrm{crit}$ ratio unavoidably
depends on evolution, it is still a good proxy for the initial
rotation rates and only weakly dependent on mass, age, or
metallicity. The earliest phase, $t/t_\mathrm{MS} \leq 0.1$,
corresponds to the time taken to reach a quasi-equilibrium state,
while the late phase at $t/t_\mathrm{MS} \geq 0.9$ represents the
star's deceleration as it evolves off the MS.

Therefore, we apply a similar analysis as in
Section~\ref{sec:vdist} to the $v/v_\mathrm{crit}$ ratio. Considering
the significant $v/v_\mathrm{crit}$ variations that occur during the
early and late phases of MS evolution, we limit our analysis to stars
characterized by $0.1 \leq t/t_\mathrm{MS} \leq 0.9$, which covers
86\% of our sample. The critical velocity was adopted from
\citet{2013A&A...553A..24G} by interpolating the grid of masses and
relative ages for a given metallicity. Note that rapid rotation can
increase the MS lifetime \citep{2013A&A...553A..24G} by transporting
fresh hydrogen into the core. Depending on the initial rotation rate,
the stellar lifetime could be extended by up to 30\%. This was not
considered in estimating the ages for our sample based on the
non-rotating stellar evolution models. Nevertheless, estimates of the
relative age, which represents the fraction of a star's lifetime spent
on the MS, is not affected by this effect.

\begin{figure*}[ht!]
\plotone{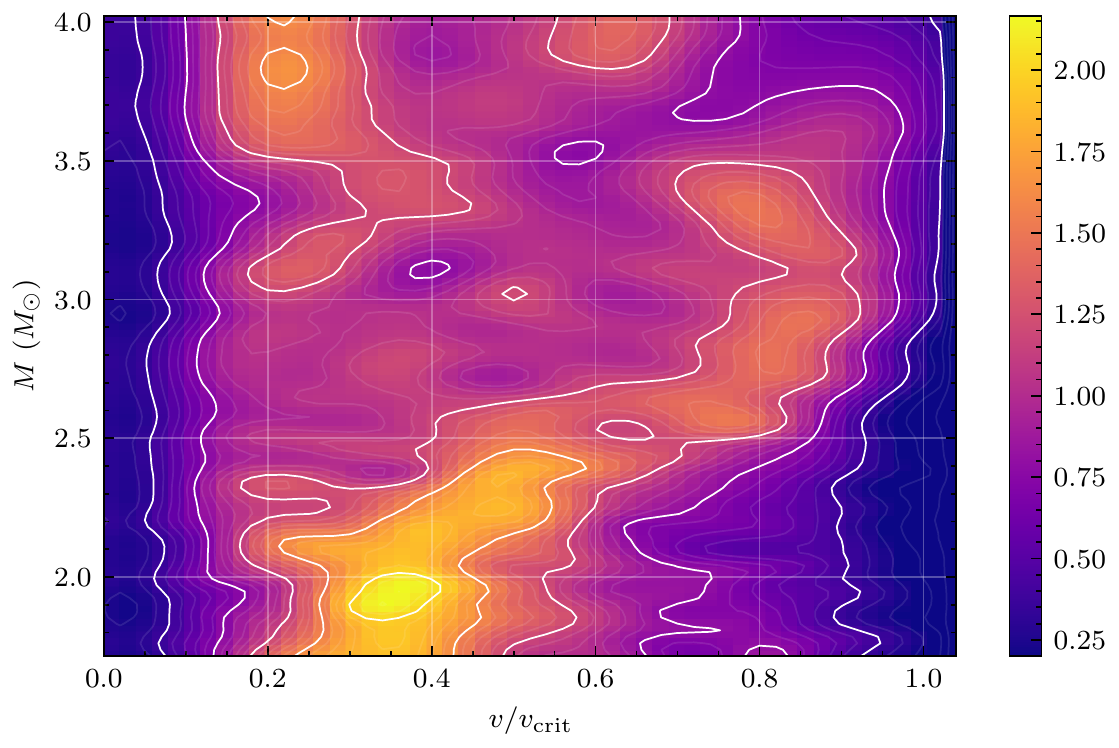}
\caption{Distribution of $v/v_\mathrm{crit}$ for normal stars as a
  function of stellar mass, color-coded by density in the
  one-dimensional normalized distribution. Iso-density contours are
  shown as white curves.\label{fig:omega_map}}
\end{figure*}

An overview of the distribution of $v/v_\mathrm{crit}$ as a function
of mass for normal stars is displayed in
Figure~\ref{fig:omega_map}. Using the same color scale as before, the
figure is color-coded by the relevant number density pertaining to the
one-dimensional normalized distribution.

\begin{figure*}[ht!]
\plotone{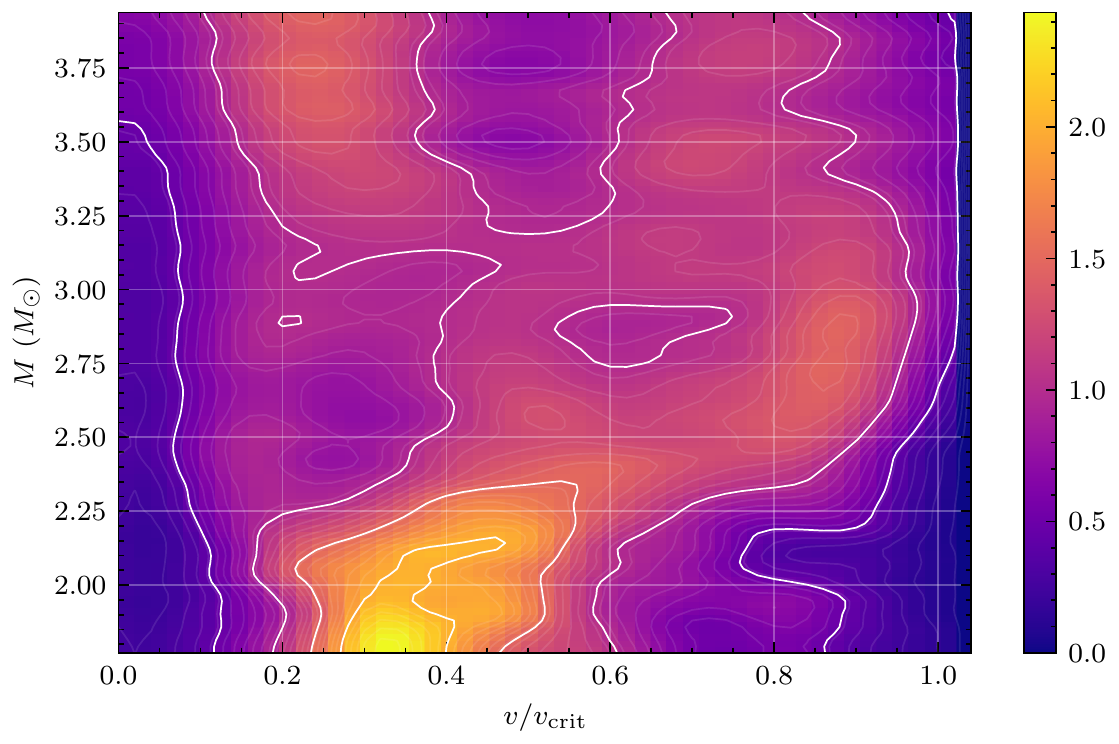}
\caption{As Figure~\ref{fig:omega_map}, but for a subsample with SNR
  $> 50$ in the \textit{g} band, overlapping with LAMOST DR5. This
  subsample reproduces the parent sample of
  \citet{2019ApJS..242...13Q}, \citet{2020A&A...640A..40H}, and
  \citet{2021A&A...645A..34P} and reduces the impact of contamination
  by CP stars by generating a cleaner sample than that used for
  Fig.~\ref{fig:omega_map}.\label{fig:omega_map_dr5}}
\end{figure*}

For less massive stars ($M<\unit[2.5]{M_\odot}$), the distribution of
$v/v_\mathrm{crit}$ is unimodal, with a significant
concentration around the peak velocity, particularly in the mass range
$\unit[1.8]{M_\odot} \leq M \leq \unit[2.0]{M_\odot}$. The velocity
ratio exhibits a clear dependence on stellar mass, in the sense that
it increases from $\sim 0.3$ to $\sim 0.55$ as mass increases, which
is consistent with \citet{2012A&A...537A.120Z}. However, a lack of
slow rotators ($v\leq \unit[100]{km\,s^{-1}}$) previously reported for
this mass range is not seen in our data. Since
\citet{2012A&A...537A.120Z} found that slow rotators are
underrepresented even when they are not removed from the sample, they
claimed that this lack is unlikely due to contamination by CP
stars. Instead, they attributed it to sampling effects in the observed
volume. However, in our sample we note that there are fewer slow
rotators following our selection of a subsample with SNR $> 50$ in the
\textit{g} band and overlapping with LAMOST DR5 to reproduce the
parent sample of \citet{2019ApJS..242...13Q},
\citet{2020A&A...640A..40H}, and \citet{2021A&A...645A..34P}. This
way, we can reduce the impact of contamination by CP stars by
generating a cleaner sample than that used for
Figure~\ref{fig:omega_map}. However, this relative absence of slow
rotators only affects stars with $v/v_\mathrm{crit} < 0.2$ ($v\leq
\unit[70]{km\,s^{-1}}$); a significant fraction of stars with
$\unit[70]{km\,s^{-1}}\leq v \leq \unit[100]{km\,s^{-1}}$
remains. Therefore, we conclude that, even if the reported relative
lack of stars is real, it must be affected by contamination by CP
stars but does not extend to $v\sim\unit[100]{km\,s^{-1}}$.

This trend between mass and $v/v_\mathrm{crit}$ continues to develop
as stars become more massive than $\unit[2.5]{M_\odot}$ and may
continue up to $\sim\unit[3.0]{M_\odot}$, where a mild turnover/upper
limit is seen. The branch of slow rotators becomes prominent for $M
>\unit[3.0]{M_\odot}$, but the bimodality is already present for $M
>\unit[2.5]{M_\odot}$. The reason why the slowly rotating branch is
not prominent in this figure is that it covers an extended
distribution across $v/v_\mathrm{crit}$. In fact,
\citet{2012A&A...537A.120Z} found that a large fraction of stars have
intermediate rotation velocities compared with the slow or fast
rotators, and it is not until $M >\unit[3.0]{M_\odot}$ that a gap
between the slowly and rapidly rotating branches becomes apparent. The
connecting structure is better visible in
Figure~\ref{fig:omega_map_dr5}, where the slowly rotating branch has
bifurcated from the rapidly rotating branch at
$M\sim\unit[2.5]{M_\odot}$. For masses greater than
$\sim\unit[3.25]{M_\odot}$, the gap between the slowly and rapidly
rotating branches is well-established.

As stellar masses increase to beyond $\sim\unit[3.6]{M_\odot}$, our
analysis suffers from poor sampling in the $\log T$ vs $\log L$
parameter space \citepalias[Figure 5]{Sun2021}. Massive MS stars in
early evolutionary phases are missing in our sample. Nevertheless, we
can make an educated guess of their properties based on the available
information. If we assume that younger stars behave similarly as their
older counterparts, it appears that more stars reside on the slowly
rotating branch compared with stars with masses in the range
$\unit[2.5]{M_\odot} \leq M \leq \unit[3.25]{M_\odot}$.

A similar bimodal distribution in rotation period has been observed
for late-type field dwarfs \citep[e.g.][]{2013MNRAS.432.1203M}, mainly
by the \textit{Kepler} and K2 missions
\citep{2010Sci...327..977B}. \citet{2021ApJ...913...70G} detected a
gap in the rotation periods for $\unit[0.57]{M_\odot} < M <
\unit[0.76]{M_\odot}$ based on a sample of $\sim$70,000 K2 targets and
attributed it to a departure from the \citet{1972ApJ...171..565S}
spin-down law. As a young star with its envelope initially decoupled
from its core loses angular momentum due to magnetic braking, it could
slow or even halt its spin-down if the core and envelope begin to
recouple at a later time. Despite the differences between our and
their samples in terms of the mass regime covered, such a mechanism
might also occur in early-type stars
\citep[e.g.][]{2012A&A...537A.120Z}.

\section{Dependence on metallicity}
\label{sec:metal}

Other than on mass, stellar rotation also depends on
  metallicity \citep[e.g.,][]{2008A&A...478..467E}. The meridional
circulation cells that carry angular momentum from the inner regions
to the surface become less efficient at lower metallicity. In
addition, stellar winds also cause angular momentum losses. Wind
strengths are greater for higher metallicities. The models of
\citet{2013A&A...553A..24G} and \citet{2015A&A...581A..15S} suggest
that normally evolving stars with low mass loss (e.g., low
  metallicities) may increase their rotation during MS evolution and
that the rotation distribution of early-type MS stars in
lower-metallicity environments might be different from that in
higher-metallicity environments. However, thus far, such differences
have not been observed conclusively. Studies of rotational
  velocities of O- and B-type stars in the LMC and the Small
  Magellanic Cloud (SMC) \citep[e.g.,][]{2004ApJ...617.1316P,
    2009ApJ...700..844P, 2019A&A...626A..50D} found little compelling
  evidence for differences with
  metallicity. \citet{2008A&A...479..541H} found that massive stars
  (with masses greater than $\unit[8]{M_\odot}$) in the SMC rotate
  faster than those in the solar neighborhood. (There is no
  significant difference between the distribution of rotation rates in
  the Galaxy and the LMC.) However, the analysis was hindered by field
  star contamination (similar to the situation in
  Section~\ref{sec:cluster}) because of the complicated composition of
  their sample, which includes target stars projected toward young
  clusters and a fraction of field stars.

To address the possible metallicity dependence, we selected metal-poor
($\mathrm{[M/H]}<\unit[-0.2]{dex}$) and metal-rich
($\mathrm{[M/H]}>\unit[0.2]{dex}$) subsamples. Each comprises more
than 5000 normal stars to ensure sufficient sample sizes for
statistical analysis. Using the same procedures as discussed in
Section~\ref{sec:rotdist}, we analyzed their rotation distributions as
a function of stellar mass.

\begin{figure*}[ht!]
\plotone{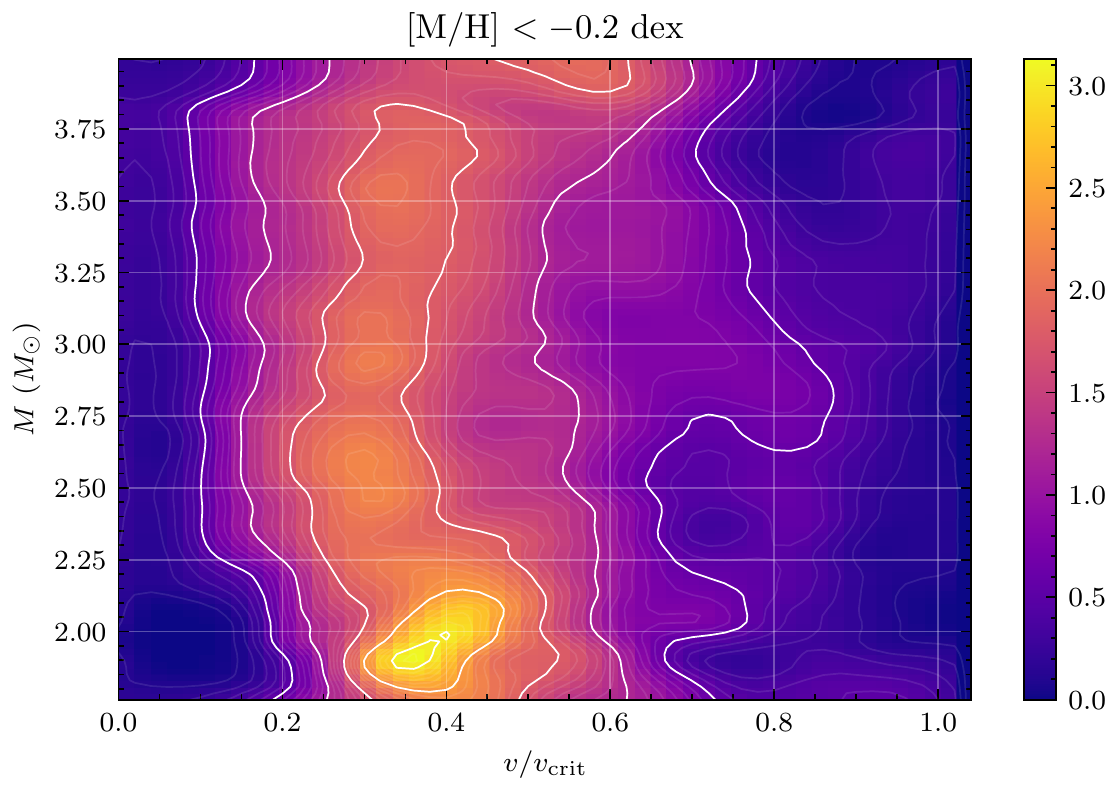}
\plotone{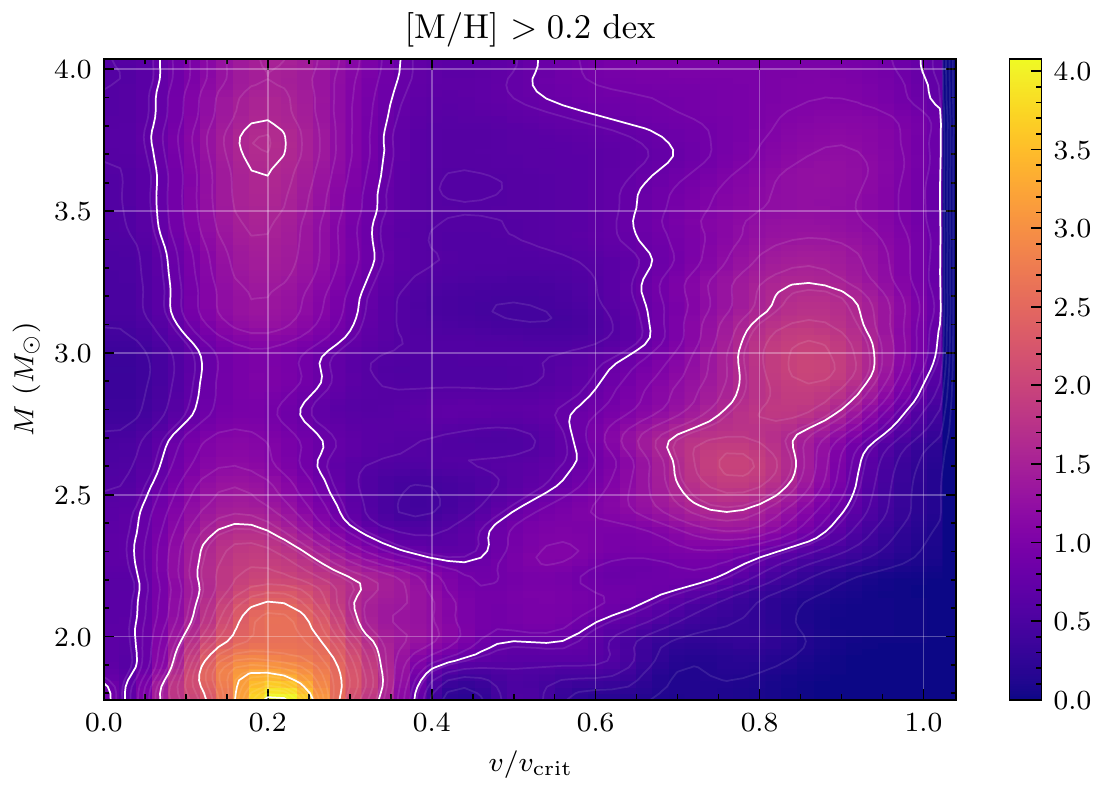}
\caption{As Figure~\ref{fig:omega_map}, but for different metallicity
  ranges. (top) Metal-poor ($\mathrm{[M/H]}<\unit[-0.2]{dex}$);
  (bottom) Metal-rich
  ($\mathrm{[M/H]}>\unit[0.2]{dex}$).\label{fig:omega_map_metal}}
\end{figure*}

The top and bottom panels of Figure~\ref{fig:omega_map_metal} show the
rotational profiles for metal-poor and metal-rich stars,
respectively. There is a significant difference between both
distributions, in the sense that metal-poor stars only exhibit a
single branch of slow rotators, with $v/v_\mathrm{crit}$ centered at
$0.3$--$0.4$, while metal-rich stars clearly show two branches of the
stellar velocity ratio. The rapidly rotating branch is similar to the
feature we saw in Figures~\ref{fig:omega_map} and
\ref{fig:omega_map_dr5}, but only for $M\geq \unit[2]{M_\odot}$. The
velocity ratio of the less massive, metal-rich stars peaks at
$v/v_\mathrm{crit}\sim 0.2$. This is much smaller than previously
observed for our entire sample. Moreover, the slowly rotating branch
is also located at lower velocity ratios than that of the metal-poor
subsample.

The shape and location of the velocity distribution pertaining to the
metal-rich subsample could be affected by slowly rotating CP
stars. Since most CP stars are metal-rich
\citep[e.g.][]{2019ApJS..242...13Q}, there is a larger fraction of
missing CP stars in this metal-rich subsample compared with both the
entire sample and the metal-poor subsample, where contamination is
expected to be minimal. \citet{2017MNRAS.468.2745N} found a linear
correlation between the masses and rotation rates of mAp stars, with
$v/v_\mathrm{crit}$ increasing from 0.08 to 0.17 as mass increases from
$\sim\unit[1.8]{M_\odot}$ to
$\sim\unit[4.5]{M_\odot}$ ($\sim 30\%$ scatter). This could be the
reason for the low rotation rate of the slowly rotating branch as seen
for the metal-rich subsample. Note that the incidence of CP stars is
not large enough to completely explain the rotation characteristics
for $M\leq\unit[2]{M_\odot}$. Thus, among less massive stars (which
are most numerous), the rotation rates of normal metal-rich stars are
lower than those for the metal-poor stars. The contribution of the
rapidly rotating branch becomes obvious for
$M\geq\unit[2.5]{M_\odot}$.

The lack of a rapidly rotating branch for the metal-poor subsample is
intriguing. One possible explanation is that this feature is
determined within the first few \unit{Myr} of a star's lifetime
through `disk locking' \citep[e.g.][]{1993A&A...272..176B}. Magnetic
interactions between a star and its accretion disk can remove angular
momentum and form intrinsically slow rotators
\citep{2011MNRAS.416..580L}, and the longer disk lifetime results in
slow rotation velocities on the MS
\citep{2020MNRAS.495.1978B}. However, \citet{2009ApJ...705...54Y,
  2010ApJ...723L.113Y} found that the near-infrared disk fraction of
low-metallicity stars declines rapidly on timescales $<\unit[1]{Myr}$,
much faster than the $\sim \unit[6]{Myr}$ observed for solar
abundance. If this were to also apply to our metal-poor subsample, and
the initial rotation rates are determined by this mechanism, the
metal-poor subsample should rotate faster than their metal-rich
counterparts. This may explain the difference in velocity ratio in the
slowly rotating branch between the metal-poor and metal-rich
subsamples, but not the absence of a rapidly rotating branch.

Another possible mechanism to slow down stellar rotation at birth is
magnetic braking \citep{2011A&A...525L..11M}. The coupling of wind
material and magnetic field lines that extend well beyond the stellar
surface exerts a torque on the stellar surface layers and thus slows
down rotation. Unlike magnetic fields in late-type stars, which are
thought to be generated through dynamo action, magnetic fields in
early-type stars are likely of fossil origin
\citep{2009ARA&A..47..333D}. \citet{2017A&A...597A..71S} proposed that
high-mass ($M\geq\unit[25]{M_\odot}$) stars develop
envelope inflation during their lifetime on the MS, which is favorably
correlated with metallicity. Therefore, stronger fossil fields are
expected for massive stars with lower metallicity than for
higher-metallicity stars under the fundamental physics law of
  flux conservation, where the surface magnetic field strength is
  expected to decrease as the stellar radius expands. If confirmed,
magnetic braking could be more prominent in the metal-poor subsample.
However, the incidence rate of magnetic fields in early-type MS
  stars is around 10\% within the solar neighborhood
  \citep{2019MNRAS.483.2300S}, and thus the lack of a rapidly rotating
  branch cannot be explained exclusively by magnetic braking.
 
\begin{figure}[ht!]
\plotone{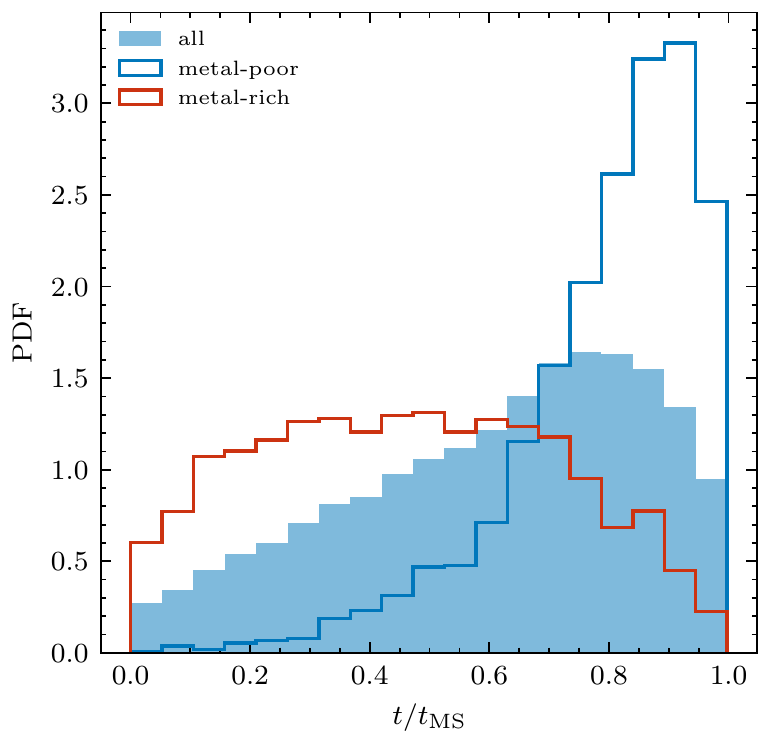}
\caption{Histograms of the relative ages of (cyan) all normal stars,
  (blue) the metal-poor subsample
  ($\mathrm{[M/H]}<\unit[-0.2]{dex}$), and (red) the metal-rich
  subsample ($\mathrm{[M/H]}>\unit[0.2]{dex}$). \label{fig:age}}
\end{figure}
 
Instead, the lack of a rapidly rotating branch could also be related
to the fact that the metal-poor subsample is generally much older than
the full sample (see Figure~\ref{fig:age}). If the overall spin down
of metal-poor stars is stronger than predicted by stellar evolutionary
models and more (specific) angular momentum is lost during the MS
phase, the rapidly rotating branch of the older metal-poor subsample
may have been moderated and may even have merged with the slowly
rotating branch. This would make it appear as a single population with
slightly higher rotation rates than characteristic of the slowly
rotating branch.

\section{Conclusions}
\label{sec:conclusions}

Using the catalog of \citetalias{Sun2021}, in this paper we have
studied the statistical characteristics of stellar rotation in the MS
evolutionary phase. The distribution of the equatorial rotational
velocities has been estimated by rectifying the error distribution and
correcting for projection effects. We also introduced the ratio
$v/v_\mathrm{crit}$ to alleviate the dependence of the rotational
velocity on mass and age.

Our main results are the following:

\begin{itemize}
\item The velocities of the less massive stars
  ($M<\unit[2.5]{M_\odot}$) exhibit a unimodal distribution, with
  their peak velocity ratio increasing from 0.3 to 0.5 as stellar mass
  increases.
\item There is no clear sign of a lack of slow rotators for
  $M\sim\unit[1.9]{M_\odot}$. Even if contamination by CP stars is
  better corrected for, the gap is smaller than previously reported.
\item A bimodal rotation distribution emerges as stars increase in
  mass ($M>\unit[2.5]{M_\odot}$). The gap between slow and fast
  rotation becomes prominent for $M>\unit[3.0]{M_\odot}$. Compared
  with the less massive stars, the rapidly rotating branch exhibits a
  milder increase with stellar mass and a turnover at
  $M\sim\unit[3.0]{M_\odot}$.
\item For stars more massive than $\unit[3.6]{M_\odot}$, more stars
  reside on the slowly rotating branch compared with stars with masses
  in the range $\unit[2.5]{M_\odot}\leq M \leq \unit[3.25]{M_\odot}$.
\item Metal-poor ([M/H] $< -0.2$ dex) stars only exhibit a single
  branch of slow rotators, with $v/v_\mathrm{crit}$ centered at
  0.3--0.4, while metal-rich ([M/H] $> 0.2$ dex) stars clearly show
  two branches. The difference could be attributed to unexpectedly
  high spin-down rates in the metal-poor subsample. Strong
    magnetic fields associated with stars with low metallicity could
    partly explain the result.
\item The bulk of the metal-poor stars are characterized by higher
  rotation rates than the metal-rich stars.
\item Roughly 10\% of CP (Am and mAp) stars exhibit abnormally high
  rotational velocities.
\end{itemize}

Our understanding of early-type stellar rotation could be improved in
two ways: (1) currently, sample decontamination largely depends on
external catalogs, whose homogeneity and completeness are difficult to
assess. This caveat is partially caused by the insufficiently high
SNR, limited number of observation epochs, and narrow wavelength
coverage of our current spectroscopic survey. In the future, armed
with more observations collected by LAMOST-MRS, combined with
LAMOST-LRS, we can improve the catalog and apply a more consistent
decontamination approach. (2) The description of stellar rotation
($v/v_\mathrm{crit}$) still depends on the stellar evolutionary models
adopted. Our sample's homogeneity and size allow us to quantitatively
study the rotation distribution as a function of stellar mass, age,
and metallicity. For instance, a variate-controlled sample could be
used to test the characteristics of stellar evolution on the MS, which
in turn could help refine and determine the free parameters remaining
in stellar rotation models. Further detailed analysis of the present
stellar sample and models of rotating stars will be pursued in
forthcoming papers.

\acknowledgments L. D. acknowledges research support from the National
Natural Science Foundation of China through grants 11633005, 11473037,
and U1631102. The Guoshoujing Telescope (the Large Sky Area
Multi-Object Fiber Spectroscopic Telescope; LAMOST) is a National
Major Scientific Project built by the Chinese Academy of
Sciences. Funding for the project has been provided by the National
Development and Reform Commission, China. LAMOST is operated and
managed by the National Astronomical Observatories, Chinese Academy of
Sciences. This work has made use of data from the European Space
Agency (ESA) mission \textit{Gaia}
(\url{https://www.cosmos.esa.int/gaia}), processed by the
\textit{Gaia} Data Processing and Analysis Consortium (DPAC;
\url{https://www.cosmos.esa.int/web/gaia/dpac/consortium}). Funding
for the DPAC has been provided by national institutions, in particular
the institutions participating in the \textit{Gaia} Multilateral
Agreement.

\vspace{5mm}
\facilities{LAMOST}
\software{PARSEC \citep[1.2S;][]{2012MNRAS.427..127B}, Astropy
  \citep{2013A&A...558A..33A}, Matplotlib \citep{2007CSE.....9...90H},
  TOPCAT \citep{2005ASPC..347...29T}}

\end{document}